%% file: Conf-SchemaEvolMapArxiv.tex
\documentclass{llncs}

\newcommand \com[1]{}

\usepackage{amsfonts}
\usepackage{graphicx,url}
\usepackage{graphics}
\usepackage{amssymb,amsmath,latexsym}

\usepackage{algorithm,algorithmic}
\renewcommand{\algorithmicrequire}{\textbf{Input:}}
\renewcommand{\algorithmicensure}{\textbf{Output:}}
\renewcommand{\algorithmicforall}{\textbf{for each}}
\renewcommand{\algorithmiccomment}[1]{\small \% \textit{#1}}

\usepackage{pst-all}
\psset{levelsep=5mm,nodesep=2.1mm,treesep=3mm,linewidth=0.15mm,linecolor=black}
\newcommand{\tree}[2]{\raisebox{.6ex}{$\pstree{\Tr{#1}}{#2}$}}

\newcommand{\leaf}[1]{\Tr{#1}}

\com{
\newtheorem{theorem}{Theorem}[section]

\newtheorem{algorithm}[theorem]{Algorithm}
\newtheorem{definition}[theorem]{Definition}
}

\newcommand{\N}{\mbox{I$\!$N}}

\def\D{\hbox{${\cal D}$}}
\def\M{\hbox{${\cal M}$}}

\newcommand{\dc}{\ifmmode \mathit{XMLCorrector}\else \textit{XMLCorrector}\fi}
\newcommand{\esch}{\ifmmode \mathit{ExtSchemaGenerator}\else \textit{ExtSchemaGenerator}\fi}
\newcommand{\xfdT}{\ifmmode \mathit{XFDTools}\else \textit{XFDTools }\fi}
\newcommand{\mapG}{\ifmmode \mathit{MappingGen}\else \textit{MappingGen}\fi}
\newcommand{\docT}{\ifmmode \mathit{XTraM}\else \textit{XTraM}\fi}

\newcommand{\ie}{\ifmmode \mathit{i.e.}\else \textit{i.e.}\fi}
\newcommand{\wrt}{\ifmmode \mathit{w.r.t.}\else \textit{w.r.t.}\fi}
\newcommand{\eg}{\ifmmode \mathit{e.g.}\else \textit{e.g.}\fi}

\newcommand{\ed}[1]{\texttt{#1}}
\newcommand{\edSetStart}{\ed{set\_startelm}}
\newcommand{\edUnsetStart}{\ed{unset\_startelm}}
\newcommand{\edInsElm}{\ed{ins\_elm}}
\newcommand{\edDelElm}{\ed{del\_elm}}
\newcommand{\edRelRoot}{\ed{rel\_root}}
\newcommand{\edRelElm}{\ed{rel\_elm}}
\newcommand{\edInsOpr}{\ed{ins\_opr}}
\newcommand{\edDelOpr}{\ed{del\_opr}}
\newcommand{\edRelOpr}{\ed{rel\_opr}}
\newcommand{\edInsRule}{\ed{ins\_rule}}
\newcommand{\edDelRule}{\ed{del\_rule}}
\newcommand{\edInsTree}{\ed{ins\_tree}}
\newcommand{\edDelTree}{\ed{del\_tree}}
\newcommand{\edInsTreeRule}{\ed{ins\_treerule}}
\newcommand{\edDelTreeRule}{\ed{del\_treerule}}

\usepackage{authblk}

\title{A ToolBox for  Conservative XML Schema Evolution and Document Adaptation}

\author{Joshua Amavi, Jacques Chabin, Mirian {Halfeld Ferrari}, Pierre R\'ety
}
\authorrunning{J. Amavi, J. Chabin, M. {Halfeld Ferrari} and P. R\'ety}
\institute{Univ. Orl\'eans, INSA Centre Val de Loire, LIFO EA 4022, FR-45067 Orl\'eans, France
\email{\small\{joshua.amavi, jacques.chabin, mirian, pierre.rety\}@univ-orleans.fr} 
}

\date{}

\begin{document}

\maketitle

\begin{abstract}
This paper proposes a set of tools to help dealing with XML database evolution.
It aims at establishing a multi-system  environment where a global integrated system  works in harmony with some local original ones,
allowing data  translation in both directions and, thus, activities on both levels.
To deal with schemas, we propose an algorithm that computes a  mapping  capable of obtaining a global schema which  is  a  conservative extension of original local schemas. 
The role of the obtained mapping is then twofold:
it ensures  schema evolution, via composition and inversion, and
it guides the construction of a document translator, allowing automatic data adaptation \wrt\ type evolution.
This paper applies, extends  and put together some of our previous contributions.





\end{abstract}

\section{Introduction}
\label{SchEvolMap-intro}
\input{SchEvolMap-intro}


\paragraph{\textbf{Motivating Example.}}
\label{SchEvolMap-mex}
\input{SchEvolMap-mex}

\paragraph{\textbf{Tools for Supporting Schema Evolution.}}
\label{SchEvolMap-tools}
\input{SchEvolMap-tools}


\section{Background}
\label{SchEvolMap-back}
\input{SchEvolMap-back}

\section{Schema Evolution}
\label{SchEvolMap-schemaEvol}

\subsection{Conservative XML Type Extension (\esch)}
\label{SchEvolMap-schAlgo}
\input{SchEvolMap-schAlgo}

\subsection{Schema Mappings }
\input{SchEvolMap-mapDef}

\subsection{Edit operations}
\input{SchEvolMap-editOp}

\subsection{Generating a Schema Mapping (\mapG)}
\label{SchEvolMap-generateMap}
\input{SchEvolMap-generateMap}

\subsection{Going Further with Mappings to support Schema Evolution}
\label{SchEvolMap-compInv}
\input{SchEvolMap-compInv}


\section{Adapting XML Documents to a New Type}

\subsection{Correcting XML Documents  (\dc)}
\input{SchEvolMap-corrector}

\subsection{Document Translation guided by  Mapping (\docT)}
\label{SchEvolMap-adaptDoc}
\input{SchEvolMap-adaptDocument}

\section{Related  Work and Concluding Remarks}
\input{SchEvolMap-relwork}

\vspace{0.1cm}
\input{SchEvolMap-FinalRem}

\bibliographystyle{llncs2e/splncs03}      

\end{document}

%% file: SchEvolMap-intro.tex
The construction of new applications aiming at integrating data from different sources while still allowing the use
of  original local systems is not an easy task.
The idea here is to establish a multi-system environment composed by a global  central system which 
 is  a \textit{conservative}  evolution of local ones, capable of processing  changes that can then be  transmitted to local systems.
The communication should be possible in both directions: local-to-global and global-to-local.
The goal is to allow independent local services to continue working on their own data, with their own tools while permitting 
  diagnosis and changes based on a general and complete view of all services.
This scenario requires tools for dealing with type evolution and  document  adaptation.
It can be useful  as a temporary configuration, deferring complete integration until local systems are ready, or as a flexible architecture adopted by the enterprise.

In this context, we suppose that  $S_1, \dots, S_n$ are local systems which deal with sets of XML documents $X_1, \dots, X_n$, respectively,
and that inter-operate with a global, integrated system $S$.
Each set $X_i$ conforms to schema or type constraints $\D_i$, while  $\D$ is an extended  type (of $S$) that accepts any local document from $\D_i$.
We assume that the global system $S$ may  evolve to  $S'$,  accepting more documents or rejecting some original ones.
Our goal is to propose tools allowing  automatic type transformation accompanied by  automatic document translation.

\vspace{0.1cm}
The contribution of this paper can be summarized as follows:

\vspace{-0.2cm}
\begin{trivlist}
\item[$\bullet$] We apply, extend and  put together some of our previous work dealing with XML constraints in the context of  database integration or evolution.
\item[$\bullet$]   In~\cite{CHMR13} we proposed \esch,  an  algorithm for generating  a new type which was  the closest conservative evolution of some given types. 
 Here,  we extend that work by generating  \textit{mappings} that indicate how to transform original schemas into the extended one (and vice-versa), via edit operations. 
 These mappings are  then used to ensure different forms of type evolution.
\item[$\bullet$]   In~\cite{ABS13} we  developed \dc\  to correct  XML documents  \wrt\  types.
We use  it here  inside  a  document translator which is guided by a schema mapping.

\end{trivlist}

\vspace{-0.2cm}
The following motivating example  illustrates the advantages of our proposal.
We then offer an overview of our realisations and  goals.

%% file: SchEvolMap-mex.tex
 We consider the hospital data  maintained by 
three services: a service which has information about patients and their treatments, another service
that is responsible for bills and  one service that keeps contact to insurance companies and  tells whether a treatment is covered by the insurance of a patient.
Figure~\ref{fig:hospitalSchema} shows a summarized version of the DTD of each service.
Notice that  we omit the definition of elements whose type is PCDATA.

\vspace{-0.2cm}
\begin{figure}[!ht]
\scriptsize
\begin{multicols}{2}
\begin{verbatim}
Patient and Treatment Service
<!ELEMENT hospital  (info*)>
<!ELEMENT info      (patient|treatment)>
<!ELEMENT patient   (SSN,pname,visitInfo*)>
\end{verbatim}
\columnbreak 
\begin{verbatim}

<!ELEMENT visitInfo (trId,date)>
<!ELEMENT treatment (trId,tname,procedure)>
<!ELEMENT procedure (treatment*)>
\end{verbatim}

\end{multicols}

\vspace*{-0.8cm}
\begin{multicols}{2}
\begin{verbatim}
Insurance coverage Service
<!ELEMENT hospital (info*)>
<!ELEMENT info     (cover|policy)>
<!ELEMENT cover    (SSN,plname)>
<!ELEMENT policy   (plname,trId*)>
\end{verbatim}

\columnbreak
\begin{verbatim}
Bill Service
<!ELEMENT hospital (info*)>
<!ELEMENT info     (bill)>
<!ELEMENT bill     (SSN,item*,date)>
<!ELEMENT item     (trId,price)>
\end{verbatim}
\end{multicols}
\vspace{-0.8cm}
\caption{DTD of the services of a hospital
\label{fig:hospitalSchema}}
\end{figure}
\vspace{-0.2cm}

Without interfering with these local services, possibly demanding to keep their own local systems,
the hospital direction may want to have a global view of all services to process reports and statistics as, for instance, the percentage of  insurance companies covering radiotherapy or myopia surgery; the number of patient paying treatments by their own, etc. 
The global system may receive information directly: a doctor having access to the global schema may introduce information about a new treatment (the price he fixed for this treatment) together with the first patients he is going to treat. 
Moreover, by analysing all  global data, the direction may decide to change its politics. For example, it should decide to introduce a discount for patients not been covered at all, provoking a schema modification (or evolution).
 
This flexibility can be reached by  permitting   some basic actions.
Firstly, the construction  of a global conservative schema capable of accepting any local document together with new documents submitted directed to the global schema. Notice that  the  local data does not need  to be translated, since it is valid \wrt\ the global schema.
Secondly, the translation of  documents from the global to a local system, allowing  updates made on the global level to be passed to the local level.
Finally, the evolution of the global schema keeping available a mapping to translate local source schema to the new global one. 
Similar actions are allowed to deal with local schema evolution.

 Figure~\ref{fig:GlobalhospitalSchema} shows   the global DTD
 resulting from the method we have proposed in~\cite{CHMR13}. 
In this paper, we introduce  an algorithm that  builds  mappings from the original DTD to the global one (built according to the ideas in  \cite{CHMR13}).
 This mapping is expressed in terms of  a sequence of edit operations.
Knowing the mapping that transforms a local schema to the global one, it is straightforward to obtain its inverse.
The inverse mapping will be the basis for translating documents from a global schema to a local one.
 
 \vspace{-0.2cm}
 \begin{figure}
 \scriptsize
 \begin{verbatim}
<!ELEMENT hospital  (info*)>
<!ELEMENT info      ((patient|treatment)|(cover|policy)|bill)>
\end{verbatim}
\vspace*{-0.3cm}
\begin{tabular}{ll}
\verb|<!ELEMENT patient   (SSN,pname,visitInfo*)>| ~~~~~~~& \verb|<!ELEMENT cover    (SSN,plname)>| \\
\verb|<!ELEMENT visitInfo (trId,date)>| & \verb|<!ELEMENT policy   (plname,trId*)>| \\
\verb|<!ELEMENT treatment (trId,tname,procedure)>| & \verb|<!ELEMENT bill     (SSN,item*,date)>| \\
\verb|<!ELEMENT procedure (treatment*)>| & \verb|<!ELEMENT item     (trId,price)>| \\
\end{tabular}

 \caption{Global DTD for the Hospital
\label{fig:GlobalhospitalSchema}}
 \end{figure}
  \vspace{-0.2cm}
 
\begin{figure}
\includegraphics[width=1.0\textwidth]{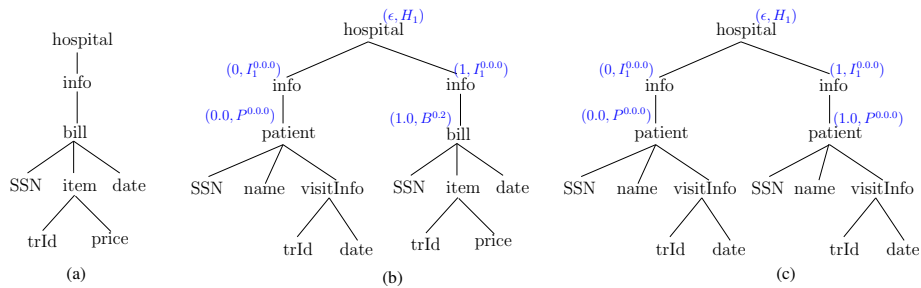}
 \caption{(a) An XML tree  valid \wrt\ the billing local schema.
 (b) An XML tree valid \wrt\ the global schema  of Figure~\ref{fig:GlobalhospitalSchema}.
 (c) Tree resulting from the translation of (b) into the patient local schema of Figure~\ref{fig:hospitalSchema}.
 Trees (b) and (c) are annotated (c.f. Section~\ref{SchEvolMap-adaptDoc})
\label{fig:Docs}}
 \end{figure}

 In Figure~\ref{fig:Docs}(a) we find a document valid \wrt\ the billing local schema. Notice that it is also valid \wrt\ the global schema of  Figure~\ref{fig:GlobalhospitalSchema}.  Figure~\ref{fig:Docs}(b) shows an XML document concerning patients and bills. 
 This document is valid \wrt\ the global schema but not valid \wrt\ to  any local schemas.
 Translating the document of  Figure~\ref{fig:Docs}(b) into a document respecting the patient schema we obtain the document of Figure~\ref{fig:Docs}(c).
The given translation is guided by the schema mapping from the global schema of Figure~\ref{fig:GlobalhospitalSchema} to the local patient schema of Figure~\ref{fig:hospitalSchema}.

\vspace{-0.2cm}
\vspace{-0.4cm}

%% file: SchEvolMap-tools.tex
We propose a set of tools to help dealing with XML database  evolution. 
Our goal is to  implement a platform where all our proposed tools will be available.
Below we describe some important modules of our \textit{ToolBox}, distinguishing   those that have been proposed
and implemented previously and those that we introduce in the current paper. 
 
 \vspace{0.1cm}

\noindent
\textsc{Tools from previous work:}\footnote{\esch\  is available on \url{http://www.univ-orleans.fr/lifo/Members/rety/logiciels/RTGalgorithms.html}.
\dc\ is available on \url{http://www.info.univ-tours.fr/~savary/English/xmlcorrector.html}.}
\vspace{-0.2cm}
\begin{trivlist}
\item[$\bullet$]   \esch (\cite{CHMR13}) extends a given 
schema $G$, seen as a regular tree grammar,  into a new grammar $G'$ respecting the following  
property: the language generated by  $G'$ is the smallest set of unranked trees that contains 
the language generated by $G$ and the grammar $G'$ is a Local Tree Grammar (LTG) or a Single-Type Tree Grammar (STTG).

\noindent

\vspace{0.1cm}
\item[$\bullet$]  \dc (\cite{ABS13})  corrects  an XML document \wrt\ schema constraints expressed as a DTD (or an LTG). 
The corrector reads the entire XML document (or tree) $t$ in order to propose solutions. 
\dc\ finds all solutions within a given threshold $th$.


\end{trivlist}

The above tools are the fundamental bricks for the new ones proposed in this paper.
The ideas introduced by \esch\ are followed in order to build our mapping generator while 
\dc\ is called by our translation module to correct parts of an XML document.
The new proposed methods are essential tools to allow schema evolution and compatible document translation.

 \vspace{0.1cm}
\noindent
\textsc{Tools introduced in the current work:}

\vspace{-0.2cm}
\begin{trivlist}
\item[$\bullet$]  \mapG:  We propose an algorithm that applies the ideas of~\cite{CHMR13} to generate a mapping from one schema $G$, seen as a regular tree grammar, to an extended schema $G'$ which will be an LTG. The resulting  schema mapping $m$  is a sequence of operations on grammar rules that indicates, step by step, how to transform $G$ into $G'$ following the approach in~\cite{CHMR13}.  Given a mapping  $m$ we can easily compute its inverse $m^{-1}$ or compose it to other mappings; allowing schemas to evolve.

\item[$\bullet$] \docT: Based on a given mapping $m$ (from schema $S$ to $T$),  we propose a method to translate an XML document  (or tree) $t$, valid \wrt\ $S$ into a document $t'$ valid \wrt\ $T$.
The edit distance between $t$  and $t'$ is no higher than  a given positive threshold  $th$.  
Moreover,  $t'$ is the closest tree to  $t$, obtained by changing $t$ according to the schema modifications imposed by $m$.
 For each edit operation on $S$, to obtain $T$,  we analyse what should be the corresponding update on  document $t$.
When this update violates validity, we use \dc\ to propose corrections to the subtree involved in the update.

\end{trivlist}

%% file: SchEvolMap-back.tex
An XML document is an {\em unranked tree}, defined in the usual way as 
a mapping $t$ from a set of positions $Pos(t)$ to an alphabet $\Sigma$.
The set of the trees over $\Sigma$ is denoted by $T_\Sigma$.
For $v \in Pos(t)$, $t(v)$ is the label of $t$ at the position $v$.
{\em Positions} are sequences of integers in $\N^*$ and $Pos(t)$ satisfies:
$\forall u,i,j\, (j \geq 0 , u.j \in Pos(t), 0 \leq i  \leq  j) \Rightarrow u.i \in Pos(t)$
(char ``.'' denotes the concatenation).
The {\em size} of $t$ (denoted $|t|$) is the cardinal of $Pos(t)$.
As usual, $\epsilon$ denotes the empty sequence of integers, i.e.\ the root position and
$t$, $t'$ will denote trees.

 Given a tree  $t$, we  denote by $t|_{p}$ the subtree whose root is at position $p \in Pos(t)$, \ie\
$Pos(t|_{p})= \{ s \mid p.s \in Pos(t) \}$ and for each $s \in Pos(t|_{p})$ we have $t|_{p}(s) = t(p.s)$.
Now, let $p \in Pos(t)$ and  $t'$ be a tree, we note $t[p \leftarrow t']$
as the tree that results of substituting the subtree of $t$ at position $p$ by $t'$.

\vspace{-0.15cm}

\begin{definition}[Regular Tree Grammar, derivation]
{\rm
A \textit{regular tree grammar} (RTG) is a 4-tuple $G = (N, \Sigma,S, P)$, where:
$N$ is a finite set of \textit{non-terminal symbols};
$\Sigma$ is a finite set of \textit{terminal symbols};
$S$ is a set of \textit{start symbols}, where $S \subseteq N$ and 
$P$ is a finite set of \textit{production rules} of the form $X \to a\, [R]$, where $X \in N$, $a \in \Sigma$, and $R$ is a regular expression over $N$.
We say that, for a production rule, $X$ is the left-hand side, $a\, [R]$ is the right-hand side, and $R$ 
is the content model.

\noindent
For an RTG $G = (N, \Sigma,S, P)$, we say that a tree $t$ built on $N \cup \Sigma$ derives (in one step) into $t'$ iff 
$(i)$ there exists a position $p$ of $t$ such that $t|_p = A \in N$ and a production rule $A \to a\, [R]$ in $P$, and 
$(ii)$ $t' = t[p \leftarrow a(w)]$ where $w\in L(R)$ ($L(R)$ is the set of words of non-terminals generated by $R$).
We write $t \rightarrow_{[p, A \to a\, [R]]} t'$.
More generally, a derivation (in several steps) is a (possibly empty) sequence of
one-step derivations. We write $t \rightarrow^*_G t'$.

\noindent
The \textit{language} $L(G)$ generated by $G$ is the set of trees containing only terminal symbols, defined by\,:
$L(G)= \{t \mid \exists A \in S,\, A \rightarrow^*_G t\}$.~\hfill{$\Box$} 
}
\end{definition}

\vspace{-0.1cm}
\noindent
\textbf{Remark:} As usual, in this paper, our algorithms
start from grammars in reduced form and (as in \cite{MaL02}) in normal form.
A \textit{regular tree grammar} (RTG) is said to be in {\bf reduced form} if
$(i)$ every non-terminal is reachable from a start symbol, and
$(ii)$ every non-terminal generates at least one tree containing only terminal symbols.
A  regular tree grammar (RTG)   is said
to be in {\bf normal form} if
distinct production rules have distinct left-hand-sides.~\hfill{$\Box$}

Among RTG  we are particularly interested in local tree grammars
which have the same expressive power as DTD\footnote{Note that converting an LTG into normal 
form produces an LTG as well.}. We recall the definition from~\cite{MLM03}:

\begin{definition}[Local Tree Grammar]
{\rm
Two non-terminals $A$ and $B$ 
(of the same grammar $G$) are said to be \textit{competing with each other} if $A \neq B$ and
$G$ contains production rules of the form $A \rightarrow a[R]$ and $B \rightarrow a[R']$ 
(i.e.\ $A$ and $B$ generate the same terminal symbol).
A \textit{local tree grammar} (LTG) is a regular tree grammar
that does not have competing non-terminals\footnote{In contrast, a \textit{single-type tree grammar} (STTG) is an RTG in normal form,
where $(i)$ for each production rule, non terminals in its regular expression do not
compete with each other, and $(ii)$ start symbols do not compete with each other. }.
A \textit{local tree language} (LTL) is a language that can be generated by at least one LTG.~\hfill{$\Box$} 
}
\end{definition}

%% file: SchEvolMap-schAlgo.tex
In~\cite{CHMR13} we find   conservative evolution algorithms  that compute a local or single-type grammar  which extends minimally a given original  
regular grammar. That paper proves the correctness and the minimality of the generated  grammars. 
In the current paper we will only deal with the generation of LTG.
We follow the idea of \esch\ which is very simple when dealing with the generation of an LTG from an RTG:
replace each pair of competing non-terminals by a new non-terminal, until there are no more competing non-terminals.
The regular expression of a new non-terminal rule is the disjunction of the  regular expressions associated to  competing non-terminals.

Let us consider the example of Section~\ref{SchEvolMap-mex} where we have  three hospital services, each one having its own
LTG (or DTD) as schema. 
Figure~\ref{fig:hospRTG} shows the RTG obtained by the union of the production rules of all these three grammars while 
Figure~\ref{fig:hospLTG} shows the  resulting LTG. 
The obtained LTG is an extension of the original RTG since it generates \textit{all} trees generated by the original RTG and possibly others as well
(refer to example of Figure~\ref{fig:Docs}).
Clearly, the obtained grammar is also an extension of each hospital service grammar.

\vspace*{-0.5cm}
\begin{figure}[h]
\begin{center}
\begin{tabular}{lll}
\scriptsize
$H_1 \rightarrow hospital[I_1^*]$            & \scriptsize~~$H_2 \rightarrow hospital[I_2^*]$         & \scriptsize~~$H_3 \rightarrow hospital[I_3^*]$ \\  
\scriptsize
$I_1 \rightarrow info[P \mid T]$             & \scriptsize~~$I_2 \rightarrow info[C \mid Pol]$        &\scriptsize ~~$I_3 \rightarrow info[B]$ \\
\scriptsize
$P \rightarrow patient[S \cdot N \cdot V^*]$ & \scriptsize~~$C \rightarrow cover[S \cdot PN]$         &\scriptsize ~~$B \rightarrow bill[S \cdot It^* \cdot D]$ \\
\scriptsize
$V \rightarrow visitInfo[Id \cdot D]$        &\scriptsize ~~$Pol \rightarrow policy[PN \cdot Id^*]$~~ &\scriptsize ~~$It \rightarrow item[Id \cdot PZ]$ \\
\scriptsize
$T \rightarrow treatment[Id \cdot TN \cdot PR]$~~ &                                        &                                    \\
\scriptsize
$PR \rightarrow procedure[T^*]$                   &                                        &                                    \\
\end{tabular}
\vspace*{-0.5cm}
\end{center}
\caption{RTG obtained from the union of production rules of grammars.\label{fig:hospRTG}}
\end{figure}

\vspace*{-1cm}
\begin{figure}
\begin{center}
\begin{tabular}{lll}
\scriptsize$H_1 \rightarrow hospital[I_1^* \mid I_1^* \mid I_1^*]$& \hspace*{1.5cm}& \scriptsize$PR \rightarrow procedure[T^*]$\\
\scriptsize$I_1 \rightarrow info[(P \mid T) \mid (C \mid Pol) \mid B  ]$& &  \scriptsize$I_1 \rightarrow info[(P \mid T) \mid (C \mid Pol) \mid B  ]$\\
\scriptsize$P \rightarrow patient[S \cdot N \cdot V^*]$& & \scriptsize$Pol \rightarrow policy[PN \cdot Id^*]$\\
\scriptsize $V \rightarrow visitInfo[Id \cdot D]$& & \scriptsize$B \rightarrow bill[S \cdot It^* \cdot D]$\\
\scriptsize$T \rightarrow treatment[Id~\cdot~TN~\cdot~PR]$& &\scriptsize $It \rightarrow item[Id \cdot PZ]$
\end{tabular}
\vspace*{-0.2cm}
\caption{LTG obtained by algorithm in~\cite{CHMR13} from the RTG of Figure~\ref{fig:hospRTG}. \label{fig:hospLTG}}
\end{center}
\end{figure}

%

\vspace*{-1.0cm}
\normalsize

%% file: SchEvolMap-mapDef.tex
In the context of schema evolution, we say that  a \textit{source} schema (or grammar)  evolves to a \textit{target} schema. 
A \textit{schema mapping} is specified by an operation list, denoted as an \textit{edit script},  that should be performed on  source schema in order to obtain
the target schema. In this paper, we propose an algorithm that generates a mapping to translate an RTG $G$ into an LTG $G'$, following  the lines of~\cite{CHMR13} .
Our mapping is composed by a sequence of  \textit{edit operations} that should be applied on the rules of grammar $G$ in order to obtain $G'$.
Before defining all our edit operations we formally introduce the notions of edit script and  schema mapping.

In the following definition, let $ed$ be an \textit{edit operation} defined on RTG $G$. 
We denote by $ed(G)$ the RTG obtained by applying $ed$ on $G$.
Each edit operation is associated with a cost that can be fixed according to the user's priority. 
Thus, the cost of an edit script is the sum of the costs of the edit operations composing it.

\begin{definition}[Edit Script and Edit Script Cost]
\label{def:edit-script}
{\rm
An \textit{edit script} $m=\langle ed_1, ed_2, \ldots ed_n \rangle$ is a sequence of \textit{edit operations} $ed_k$ where $1 \le k \le n$.
Let $G$ be an RTG, an \textit{edit script} $m=\langle ed_1, ed_2, \ldots ed_n \rangle$ is defined on $G$ 
if and only if there exists a sequence of RTG $G_0, G_1, \ldots, G_n$ such that:
$(i)$ $G_0 = G$  and $(ii)$  $\forall\, 1 \le k \le n$, $ed_k$ is defined on $G_{k-1}$ and $ed_k(G_{k-1}) = G_k$.
Hence, we have $m(G)=G_n$.
The empty \textit{edit script} is denoted $\langle \rangle$. 
The cost of an edit script  $m$ is defined as $cost(m) = \Sigma^n_{i=1} (cost(ed_i))$. 
\hfill{$\Box$}
}
\end{definition}

\begin{definition}[Schema Mapping]
\label{def:mapping}
{\rm
A schema mapping is a triple $\M = (S, T, m)$, where $S$ is the source schema,
$T$ is the target schema, and $m$ is an \textit{edit script} that transforms $S$ into $T$ (\ie, $m(S)=T$).
We say that $\M$ is syntactically \textit{specified by}, or, \textit{expressed} by $m$.
\hfill{$\Box$}
}
\end{definition}

%% file: SchEvolMap-editOp.tex
In this section,  we define edit operations on an RTG $G = (N, \Sigma, S, P)$. 
The idea is, firstly, to represent production rules as trees. 
Then, the problem of changing one RTG into another is treated as a tree editing problem.

\vspace*{-0.2cm}
\subsubsection{Tree Representation for Production Rules.}

Let  $X \to a\, [R]$ be a production rule. We denote by $reg(X)$ the regular expression
$R$ associated with the non-terminal $X$, and by $term(X)$ the terminal symbol $a$ for the non-terminal $X$.
Note that $reg(X)$ and $term(X)$ are defined in a deterministic way since we always suppose that grammars are in normal form, therefore $X$ occurs only once as the left-hand-side of a production rule.
We treat the regular expression $R$ as an unranked tree denoted $t_R$. 
The set of non-terminal symbols occurring in $R$ is denoted by $nt(R)$.
Formally, $t_R$ is recursively defined as follows:
\begin{itemize}
  \item if $R=\epsilon$ then $t_R$ is a single node labeled by $\epsilon$.
  \item if $R=A$ where $A \in N$ then $t_R$ is a single node labeled by $A$.
  \item if $R=R_1. \cdots . R_n$ then $t_R = .(t_{R_1},\cdots,t_{R_n})$ \ie\ $t_R$ is tree such that the root is '$.$' with
        the subtrees $t_{R_1},\cdots,t_{R_n}$.
  \item if $R=R_1| \cdots | R_n$ then $t_R = |(t_{R_1},\cdots,t_{R_n})$ \ie\ $t_R$ is tree such that the root is '$|$' with
        the subtrees $t_{R_1},\cdots,t_{R_n}$.
  \item if $R=R_1^*$ then $t_R = *(t_{R_1})$.
  \item if $R=(R_1)$ then $t_R = t_{R_1}$.
\end{itemize}

We represent the right-hand side of a production rule $X \to a\, [R]$ as a tree denoted $t^r_X$ such that $t^r_X = a(t_R)$. The root of 
$t^r_X$ is the terminal $a$ which has only one subtree $t_R$. We have that $t^r_X|_0 = t_R$. 
For example, in Figure~\ref{fig:fig-nodeEdit}, the tree on the top left corner is  $t^r_I$ with $I \to info [T.(Y \mid Co)]$.

\begin{definition}[Well Formed Tree]\label{def:wellformed}
{\rm
A tree $t$ representing the right-hand side of a production rule is well formed iff the following conditions
are verified:\\
$(i)$  the root is a terminal symbol, \ie\, $t(\epsilon) \in \Sigma$, and has exactly one child;\\
$(ii)$ the leaves nodes are in $N \cup \{\epsilon\}$ and \\
$(iii)$ the internal nodes are in the set $\{|,.,*\}$ such that:
  		  if an internal node is in $\{*\}$ then the internal node has exactly one child;
  		 otherwise if an internal node is in $\{|,.\}$ then the internal node has at least one child.	\hfill{$\Box$}
}
\end{definition}

\subsubsection{Elementary Edit Operations.}
We  define  elementary edit operations by using rewriting.
Given a set of variables $X$, a rewrite rule
(written $l \rightarrow r$) is a pair of terms over $\{\mid,.,*\} \cup N \cup X$, assuming that variables have no children. 
A hedge is a (possibly empty) sequence of trees, like $[t_0,\ldots,t_n]$.
Let $h$ be a hedge, $|h|$ denotes the number of trees in $h$.
For example, if $h=[t_0,\ldots,t_n$], then $|h|=n+1$.
A substitution $\sigma$ is a mapping of finite domain from $X$ into the set of hedges, whose application is extended homomorphically to trees. 
Let $t$, $t'$ be trees, $t$ rewrites into $t'$ at position $u$, by the rewrite rule $l \rightarrow r$, and with the substitution $\sigma$
(written $t \rightarrow_{[u,l \rightarrow r,\sigma]} t'$)
if $t|_u = \sigma(l)$ and $t'=t[u \leftarrow \sigma(r)]$.
For example, given the rule $\tree{f}{\leaf{x} \leaf{y}} \rightarrow \tree{g}{\leaf{y} \leaf{x}}$ ($x$ and $y$ are variables), then
$\tree{f}{\leaf{a} \leaf{b} \leaf{c}}$ rewrites into $\tree{g}{\leaf{c} \leaf{a} \leaf{b}}$
(with substitution $x/[a,b],\,y/[c]$),
and also into $\tree{g}{\leaf{b} \leaf{c} \leaf{a}}$ (with $x/[a],\,y/[b,c]$).

In the following definition, terms are always rewritten at position $u$, with substitution $\sigma$, provided the
condition (if any) is satisfied. We only mention the rewrite rule, which is not always the same.

\begin{figure*}[!htb]
		\begin{center}
		  \includegraphics[width = 6cm]{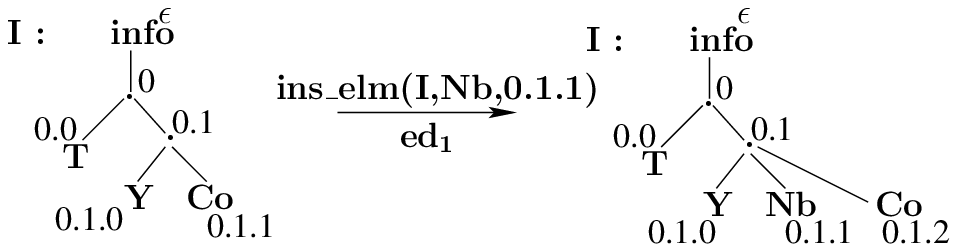}
		  \includegraphics[width = 6cm]{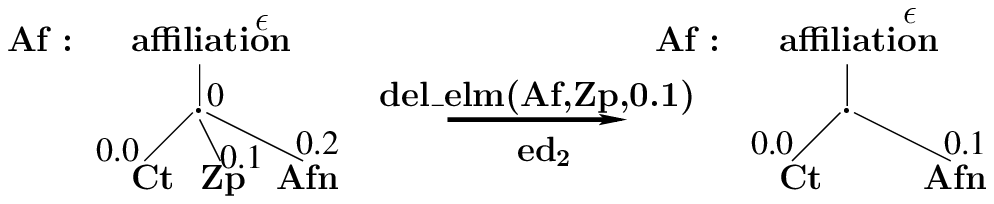}
  		  \includegraphics[width = 6cm]{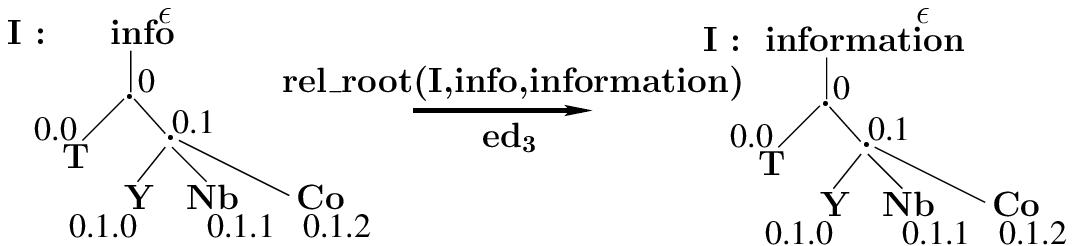}
		  \includegraphics[width = 6cm]{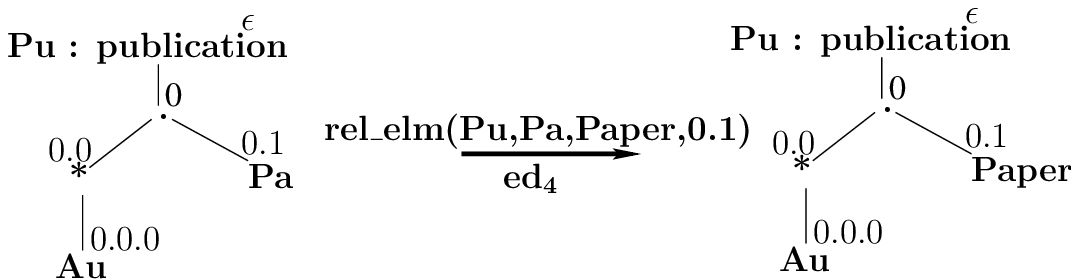}
		   \includegraphics[scale=0.6]{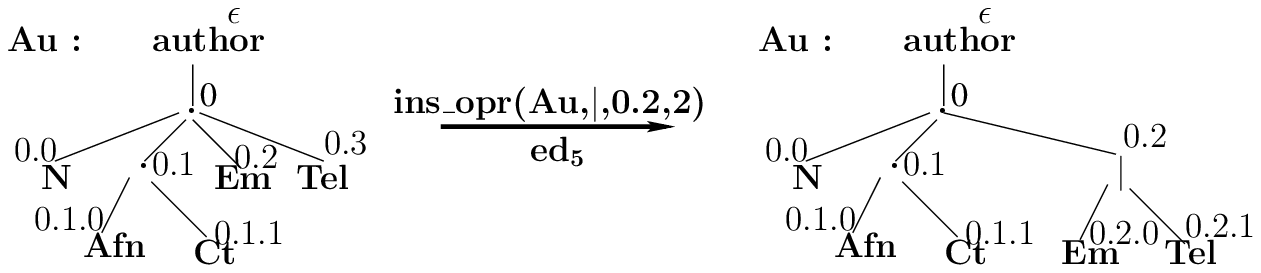}
		  \includegraphics[width = 6cm]{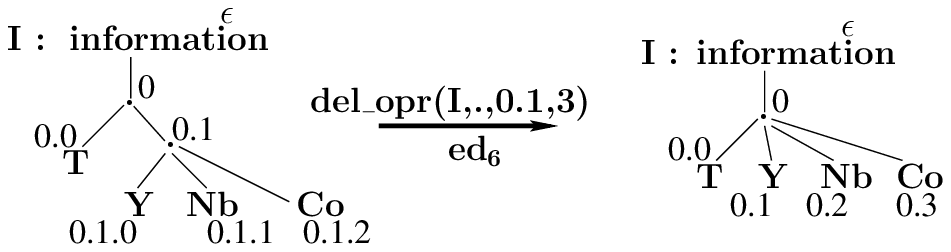}
		  \includegraphics[width = 6cm]{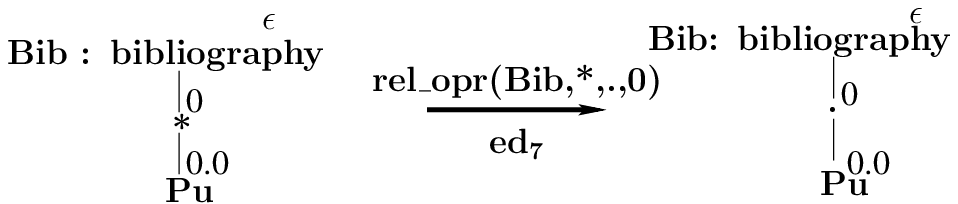} 
		\end{center}
		  \vspace{-0.5cm}
	    \caption{Example of \textit{elementary edit operations}.} 
	    \label{fig:fig-nodeEdit}
	\end{figure*}

\begin{definition}[Elementary Edit Operations]
\label{def:nodeEdit-operation}
{\rm
Given an RTG $G = (N, \Sigma, S, P)$ in normal form, an \textit{elementary edit operation} $ed$ is a partial function that transforms
$G$ into a new RTG $G'$. The \textit{elementary edit operation} $ed$ can be applied on $G$ only if $ed$ is defined on $G$.  
We distinguish four types of \textit{elementary edit operations} on RTG: 
\begin{enumerate}
  \item Edit operations to modify the set of start symbols $S$
  	    \begin{itemize}
  	      \item \edSetStart$(A)$: adds the non-terminal $A$ to $S$ where $A \in N$.
  	      \item \edUnsetStart$(A)$: deletes the non-terminal $A$ from $S$ where $A \in N$.
		 \end{itemize}
  \item Edit operations to modify non-terminal or terminal symbols in a content model
  		\begin{itemize}
  		  \item \edInsElm$(X,A,u.i)$: (cf. Figure~\ref{fig:fig-nodeEdit}($ed_1$)) applies the rewrite rule
  		  		\begin{center}
  		         \includegraphics[scale=0.8]{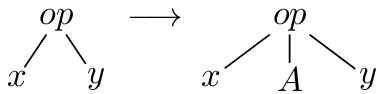}
  		        \end{center}
  		        on $t^r_X$ at position $u$ where $X \in N$, $A \in N \cup \{\epsilon\}$, $|\sigma(x)|=i$ and $op \in \{|,.\}$.

  		  \item \edDelElm$(X,A,u.i)$: (cf. Figure~\ref{fig:fig-nodeEdit}($ed_2$))  applies the rewrite rule
  		  		\begin{center}
  		         \includegraphics[scale=0.8]{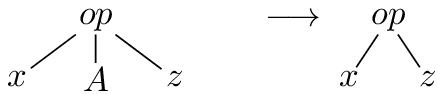}
  		        \end{center}
  		        on $t^r_X$ at position $u$ where $X \in N$, $|\sigma(x)|=i$, $A \in N \cup \{\epsilon\}$, $|\sigma(x)| + |\sigma(z)| \ge 1$ and $op \in \{|,.\}$.

  		  \item \edRelRoot$(X,a,b)$: (cf. Figure~\ref{fig:fig-nodeEdit}($ed_3$))   applies  the rewrite rule 
  		  		\begin{center}
  		         \includegraphics[scale=1]{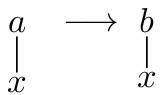}
  		        \end{center}
  		        on $t^r_X$ at position $\epsilon$ where $X \in N$, $a, b \in \Sigma$ and $|\sigma(x)|=1$.

  		  \item \edRelElm$(X,A,B,u)$: (cf. Figure~\ref{fig:fig-nodeEdit}($ed_4$)) applies the rewrite rule $A \longrightarrow B$
  		        on $t^r_X$ at position $u$ where $X \in N$, $A, B \in N \cup \{\epsilon\}$.

  		\end{itemize}
  \item Edit operations to modify operator symbols in a content model
  		\begin{itemize}
  		  \item \edInsOpr$(X,opr,u.i,n)$: (cf. Figure~\ref{fig:fig-nodeEdit}($ed_5$)) applies  the rewrite rule
  		  		\begin{center}
  		         \includegraphics[scale=0.8]{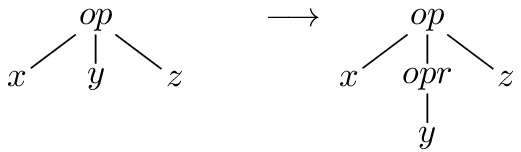}
  		        \end{center}
  		        on $t^r_X$ at position $u$ where $X \in N$, $n \ge 1$, $op \in \{|,.,*\} \cup \Sigma$, $|\sigma(x)|=i$, $|\sigma(y)|=n$ and 
                 if $n=1$ then $opr \in \{|,.,*\}$ otherwise $opr \in \{|,.\}$.

  		  \item \edDelOpr$(X,opr,u.i,n)$: (cf. Figure~\ref{fig:fig-nodeEdit}($ed_6$)) applies  the rewrite rule
  		  		\begin{center}
  		         \includegraphics[scale=0.8]{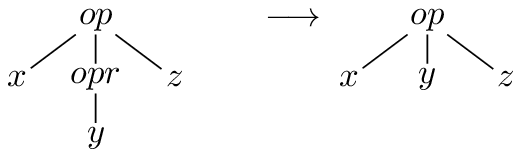}
  		        \end{center}
  		        on $t^r_X$ at position $u$ where $X \in N$, $op \in \{|,.,*\} \cup \Sigma$, $opr \in \{|,.,*\}$, $|\sigma(x)|=i$, and
  		        $|\sigma(y)|=n$. If $op \in \{*\} \cup \Sigma$ then $|\sigma(y)| = 1$ and $|\sigma(x)| + |\sigma(z)|=0$.

  		  \item \edRelOpr$(X,op,opr,u)$: (cf. Figure~\ref{fig:fig-nodeEdit}($ed_7$)) applies  the rewrite rule
  		  		\begin{center}
  		         \includegraphics[scale=1]{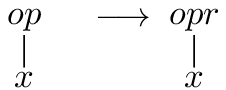}
  		        \end{center}
  		        on $t^r_X$ at position $u$ where $X \in N$, $op, opr \in \{|,.,*\}$, $|\sigma(x)| \neq 0$ and if $opr = *$ then $|\sigma(x)| = 1$.
  		  		
  		\end{itemize}  
  \item Edit operations to modify the set of production rules $P$
  		\begin{itemize}
  		  \item \edInsRule$(A,a)$: adds the new production rule $A \to a\, [\epsilon]$ to $P$ and the non-terminal $A$ to $S$,
  		        where $A \not\in N$.
  		  \item \edDelRule$(A,a)$: deletes the production rule associated with $A$ from $P$, where $A \in N$ and $reg(A) = \epsilon$. 
  		  		If $A \in S$ then $A$ is also deleted from~$S$. 
  		\end{itemize}
\end{enumerate}

\noindent
After each edit operation, the sets $\Sigma$ and $N$ are automatically updated to contain 
all and only  the terminal (resp. non-terminal) symbols appearing in  $P$.
\hfill{$\Box$}
}
\end{definition}

\begin{proposition}
An \textit{edit operation}  applied on  an  RTG $G$  results in  an  RTG $G'$ that is also in normal and reduced form.\hfill{$\Box$}
\end{proposition}

\begin{proposition}
Let $G$ and $G'$ be two RTG. There exist an edit script, composed   only  by operations of  Definition~\ref{def:nodeEdit-operation},
that  transforms $G$ into $G'$. \hfill{$\Box$}
\end{proposition}

\vspace{-0.6cm}
\subsubsection{Non-Elementary Edit Operations.}
For readability and cost estimation, we define short-cut operations, \ie, operations  seen as a one-block operation
but equivalent to a sequence of \textit{elementary edit operations}. In this paper, we introduce just those that are used
in our algorithm: 
\begin{itemize}
 \item \edInsTree$(X,R,u.i)$ (and, respectively, \edDelTree$(X,R,u.i)$): consists of inserting (respect. deleting) a subtree at a given position in the right-hand side
        of a rule. This operation is similar to \edInsElm (respect. \edDelElm) but instead of adding (respect. deleting)  a node in $t^r_X$, it adds (respect. deletes)
        the subtree $t_R$.


 \item \edInsTreeRule$(A,a,R)$ (and, respectively,  \edDelTreeRule$(A,a,R)$): adds (respect. deletes) the new (respect. existing) productive rule $A \to a\, [R]$ to $P$ and the non-terminal $A$ to $S$,   		        where $A \not\in N$ (respect.   where $A \in N$)  and $nt(R) \subseteq N$. We can transform this operation in a sequence
 that adds  $A \to a\, [\epsilon]$ to $P$ and then changes the regular expression $\epsilon$ into  $R$ 
 (respect., that changes $t_R$ into $\epsilon$ and then deletes the rule $A \to a\, [\epsilon]$ from $P$).

%
\end{itemize}

\com{
\begin{definition}[Tree Edit Operations]
\label{def:treeEdit-operation}
{\rm
Given an RTG $G = (N, \Sigma, S, P)$ in normal form, a \textit{tree edit operation} $ed$ is a partial function that transforms
$G$ into a new RTG $G'$. The \textit{tree edit operation} $ed$ can be applied on $G$ only if $ed$ is defined on $G$.  
Each \textit{tree edit operation} can be expressed as a sequence of \textit{elementary edit operations}.
\begin{enumerate}
	  \item $ins\_tree(X,R,u.i)$: consists of applying the rewrite rule
	  		\begin{center}
	         \includegraphics[scale=1]{fig-instree.eps}
	        \end{center}
	        on $t^r_X$ at position $u$ where $X \in N$, $|\sigma(R)|=1$, $|\sigma(x)|=i$ and $op \in \{|,.\}$.
	        The operation $ins\_tree(X,R,u.i)$ is equivalent to the sequence of all operations $ins\_elm(X,t_R(v),u.i.v)$, $\forall\, v \in Pos(t_R)$.
	        The insertion of these nodes are done from  root to leaves.
	        This operation is illustrated in Figure~\ref{fig:fig-treeEdit}($ed_1$).
	        
	  \item $del\_tree(X,R,u.i)$: consists of applying the rewrite rule
	  		\begin{center}
	         \includegraphics[scale=1]{fig-delelmb.eps}
	        \end{center}
	        on $t^r_X$ at position $u$ where $X \in N$, $|\sigma(x)|=i$, $|\sigma(y)|=1$ and $\sigma(y)=t_R$, $|\sigma(x)| + |\sigma(z)| \ge 1$ and $op \in \{|,.\}$.
	        The operation $del\_tree(X,R,u.i)$ is equivalent to the sequence of all operations $del\_elm(X,t_R(v),u.i.v)$, $\forall\, v \in Pos(t^r_X|_{u.i})$.
	        The deletion of these nodes are done from  leaves to root.
	        This operation is illustrated in Figure~\ref{fig:fig-treeEdit}($ed_2$).\hfill{$\Box$}
\end{enumerate}
}
\end{definition}

\begin{figure*}[!htb]
		\begin{center}
		  \includegraphics[width=6.5cm,height=2.5cm]{fig-treeEdit-1.eps}
		  \includegraphics[width = 5.5cm,height=2.5cm]{fig-treeEdit-2.eps}
		 		\end{center}
	    \caption{Example of \textit{tree edit operations}.} 
	    \label{fig:fig-treeEdit}
	\end{figure*}
}	

	Now, for each \textit{edit operation} $ed$, we define a  non-negative and application-dependent cost.
	On the one hand, we assume that  operations that do not  change the language generated by the RTG $G$ on which they were applied,  are $0$-cost.
	Their goal is just to  simplify a given regular expression. For instance,  
\edDelOpr$(X,opr,u.i)$ where $t^r_X(u) = t^r_X(u.i) = opr$ and
\edDelOpr$(X,opr,u.i)$ where $t^r_X(u.i) \in \{|,.\}$ and $t^r_X(u.i)$ has exactly one child, are 0-cost operations.
On the other hand, we suppose that an \textit{elementary edit operation} (Definition~\ref{def:nodeEdit-operation}) costs $1$,
while  a \textit{non-elementary edit operation} costs $5$. 

%% file: SchEvolMap-generateMap.tex
Algorithm~\ref{algo-rtgToltg} generates a mapping that converts an RTG in an LTG by following the ideas in~\cite{CHMR13},  explained in Section~\ref{SchEvolMap-schAlgo}.
This algorithm starts by determining a set of competing non-terminals $EC_a$ (lines~\ref{inT}-\ref{finT}).
 Then we can take arbitrarily in $EC_a$, one of these non-terminals (say $X_0$) to represent all others, \ie, when merging rules of competing terminals, one non-terminal name is chosen
 to represent the result of the merge (line~\ref{choosedNT}). 
 Recall that edit operations always deal with a production rule in its  tree-like format.
  The  new production rule of $X_0$  is built in two steps.
We add an OR operation  as the  parent of its original regular expression $reg(X_0)$ (line~\ref{addOr})
 and  then we insert   all  regular expressions associated with its  competing non-terminals as siblings of $reg(X_0)$  (line~\ref{addCompeting}).
 In line~\ref{replaceNt} we just replace,  in all production rules,  non-terminals in $EC_a$ by $X_0$.
 Original rules of non-terminals in $EC_a$  are deleted (line~\ref{deleteNt})  after, possibly, adjusting start symbols (line~\ref{adjustStart}).

\algsetup{indent=1.5em}
\begin{algorithm}[!ht]
\caption{A mapping for transforming an RTG into an LTG}
\label{algo-rtgToltg}
\begin{algorithmic}[1]
\REQUIRE A Regular Tree Grammar $G = (NT, \Sigma, S, P)$
\ENSURE An \textit{edit script} $m$ between $G$ and the LTG $G'$ such that $L(G) \subseteq L(G')$

\STATE $m := \langle \rangle$ 

\FORALL{terminal symbol $a \in \Sigma$} \label{inT}
\STATE $EC_a \!=\! \{ X_0, \ldots, X_k \}$ is a set of competing non-terminals where $term(X_i) = a$\label{finT}
\STATE Non-terminal $X_0$ is choosed to represent $X_0, \ldots, X_k$ \label{choosedNT}
\STATE Add \edInsOpr$(X_0,|,0,1)$ to $m$ \label{addOr}

\FORALL{non-terminal $X_i \in \{ X_1, \ldots, X_k \}$} \label{inEC}
\STATE Add \edInsTree$(X_0,reg(X_i),0.i)$ to $m$ \label{addCompeting}
\STATE Add \edRelElm$(Y,X_i,X_0,u)$ to $m$, for all $u$ where $u$ is the position of $X_i$ 
       \\\hspace{3.5cm}~~ in the rule $Y \to b\, [R]  \in P$ \label{replaceNt}
\STATE Add \edSetStart$(X_0)$ to $m$ where $X_0 \not\in S$ and $X_i \in S$ \label{adjustStart}
\STATE Add \edDelTreeRule$(X_i,a,reg(X_i))$ to $m$ \label{deleteNt}
\ENDFOR
\ENDFOR
\RETURN $m$
\end{algorithmic}
\end{algorithm}

\noindent
Consider the RTG of  Figure~\ref{fig:hospRTG}.
 Algorithm~\ref{algo-rtgToltg} returns the  following mapping $m$:\\
\footnotesize
$\langle \edInsOpr(H_1,|,0,1), ~\edInsTree(H_1,reg(H_2),0.1), ~\edDelTreeRule(H_2,hospital,reg(H_2)),$
$~ \edInsTree(H_1,reg(H_3),0.2), ~\edDelTreeRule(H_3,hospital,reg(H_3)), $ $\edInsOpr(I_1,|,0,1),$ \\
$\edInsTree(I_1,reg(I_2),0.1), ~ \edRelElm(H_1,I_2,I_1,0.1.0), ~ \edDelTreeRule(I_2,info,$ $reg(I_2)),$ \\
$ \edInsTree(I_1,reg(I_3),0.2), ~ \edRelElm(H_1,I_3,I_1,0.2.0), ~ \edDelTreeRule(I_3,info,reg(I_3)) \rangle. $
\normalsize
When $m$ is applied on  the RTG of Figure~\ref{fig:hospRTG}, 
the LTG of Figure~\ref{fig:hospLTG} is obtained.

\begin{proposition}
Let  $m$ be the mapping obtained by Algorithm~\ref{algo-rtgToltg} from an RTG $G$.
The language $L(m(G))$ is the least LTL that contains $L(G)$. Moreover, the grammar $m(G)$ equals
the one obtained by $\esch$. ~$\hfill{\Box}$
\end{proposition}

\vspace{-0.3cm}

\com{
Figure~\ref{fig:hospLTG} shows the LTG  resulting from the application of the following mapping over the RTG of Figure~\ref{fig:hospRTG}:
\footnotesize
$\langle ins\_opr(H_1,|,0,1), ~ins\_tree(H_1,t^r_{H_2}|_0,0.1), ~del\_tree\-rule(H_2,$ $hospital,$ $reg(H_2)),$
$~ ins\_tree(H_1,t^r_{H_3}|_0,0.2), ~del\_treerule(H_3,hospital,reg(H_3)), $ $ins\_opr(I_1,|,0,1),$
$ins\_tree(I_1,t^r_{I_2}|_0,0.1), ~ rel\_elm(H_1,I_2,I_1,0.1.0), ~ del\_treerule(I_2,info,$ $reg(I_2)),$
$ ins\_tree(I_1,t^r_{I_3}|_0,0.2), ~ rel\_elm(H_1,I_3,I_1,0.2.0), ~ del\_treerule(I_3,info,reg(I_3)) \rangle $
\normalsize
}

%% file: SchEvolMap-compInv.tex
\begin{figure*}[!htb]
\begin{center}
  \includegraphics[scale=0.85]{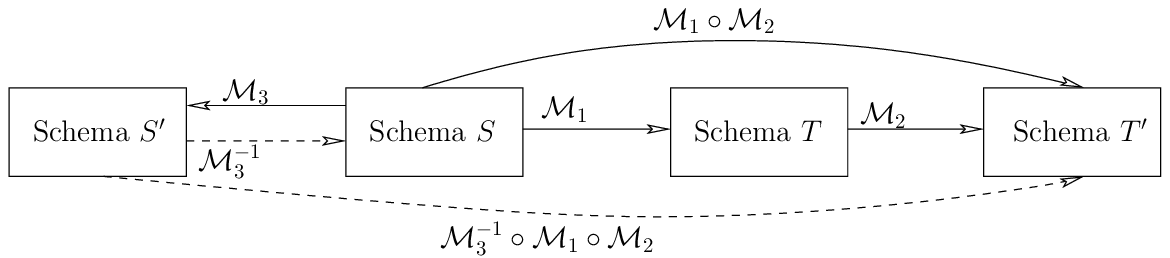} 
\end{center}
	\vspace*{-0.5cm}
  \caption{Application of composition and inversion in schema evolution.} 
  \label{fig:appCompInv}
\end{figure*}

In~\cite{FKPT11}, it was shown how two fundamental operators on schema mappings, namely composition and inversion, can be used to address
the mapping adaptation problem in the context of schema evolution. 
Given  $\M_1 =(S, T, m_1)$, a mapping between XML schemas $S$ and $T$, when $S$ or $T$ evolve, $\M_1$ shall be adapted.
By using composition and inversion operators, one can avoid mapping re-computation.
The idea is illustrated in  Figure~\ref{fig:appCompInv}.
Firstly, suppose the target schema evolves to  $T'$, and that this evolution is modeled by mapping $\M_2$.
Composing $\M_1$ and $\M_2$, denoted by $\M_1 \circ \M_2$, is an operation that has the same effect as applying first $\M_1$ and then $\M_2$.
For example, given an RTG $G$, Algorithm~\ref{algo-rtgToltg}  obtains $\M_1$ and we can find an LTG $G_1$.
If  $G_1$ evolves into $G_2$ using $\M_2$, the translation of the original $G$ into $G_2$ is obtained just by computing $\M_1 \circ \M_2$.
Now, suppose that the source schema evolves to a new source schema $S'$, modeled by  mapping $\M_3$.
To obtain schema $T'$ from $S'$, the composition  of $\M_3$ with $\M_1 \circ \M_2$ is not possible, since $\M_3$ and $\M_1 \circ \M_2$ are not consecutive. 
To apply composition we need, first, to compute the inversion of $\M_3$, denoted $\M_3^{-1}$, which ''undoes'' the effect of $\M_3$.
Once we obtain a suitable $\M_3^{-1}$, we can then apply the composition operator
        to produce $\M_3^{-1} \circ \M_1 \circ \M_2$. 
The resulting schema mapping  is now from $S'$ to $T'$. 
We now precise the notions of composition and inversion in our context.
 
 \vspace*{-0.3cm}
 \vspace*{-0.5cm}

\begin{definition}[Mapping composition and inversion]
\label{def:mappingComInv}
{\rm
Given two mappings $\M_1=(S,T,m_1)$ and $\M_2=(T,V,m_2)$, the composition of $\M_1$ and $\M_2$ 
is the mapping $\M_1 \circ \M_2 = (S,V, m_1 \,.\, m_2)$.
If  $m_1=\langle ed_1,\cdots, ed_n \rangle$, then 
the inverse of mapping $\M_1$
is the mapping $\M_1^{-1}=(T,S,m_1^{-1})$ where $m_1^{-1}=\langle ed_n^{-1},\cdots, ed_1^{-1} \rangle$ and $ed_k^{-1} (1 \le k \le n)$ is defined in 
Table~\ref{table:operationInv}.
\hfill{$\Box$}
\vspace{-0.3cm}
\begin{table}[th]
\small
\begin{tabular}{ll}
\edSetStart$(A)$~~ $\rightleftharpoons$ \edUnsetStart$(A)$ ~~~~     & \edInsOpr$(X,p,u,n)$ $\rightleftharpoons$ \edDelOpr$(X,p,u,n)$  \\
\edInsElm$(X,A,u)$~~~\, $\rightleftharpoons$ \edDelElm$(X,A,u)$   & \edRelOpr$(X,p,q,u)$ $\rightleftharpoons$ \edRelOpr$(X,q,p,u)$  \\
\edRelRoot$(X,a,b)$~~~ $\rightleftharpoons$ \edRelRoot$(X,b,a)$   & \edInsRule$(A,a)$~~~~ $\rightleftharpoons$ \edDelRule$(A,a)$ \\
\edRelElm$(X,A,B,u)$ $\rightleftharpoons$ \edRelElm$(X,B,A,u)$    & \edInsTree$(X,R,u)$ $\rightleftharpoons$ \edDelTree$(X,R,u)$ \\
\end{tabular}
\begin{tabular}{l}
\edInsTreeRule$(A,a,R)$ $\rightleftharpoons$ \edDelTreeRule$(A,a,R)$
\end{tabular}
\caption{Inverse relationship for edit operations. \label{table:operationInv}}
\end{table}
}
\end{definition}
\vspace{-1cm}

To illustrate the inverse operation,  consider the mapping generated by Algorithm~\ref{algo-rtgToltg} for
the RTG of  Figure~\ref{fig:hospRTG} (Section~\ref{SchEvolMap-generateMap}). The inverse of this mapping is:
\footnotesize
$\langle \edInsTreeRule(I_3,info,reg(I_3)), ~\edRelElm(H_1,I_1,I_3,0.2.0), ~\edDelTree(I_1,reg(I_3),0.2),  $\\
$\edInsTreeRule(I_2,info,$ $reg(I_2)), ~ \edRelElm(H_1,I_1,I_2,0.1.0), ~\edDelTree(I_1,reg(I_2),0.1),$ \\
$\edDelOpr(I_1,|,0,1), ~\edInsTreeRule(H_3,hospital,reg(H_3)), ~\edDelTree(H_1,reg(H_3),0.2),$ \\
$\edInsTreeRule(H_2,hospital,reg(H_2)), ~\edDelTree(H_1,reg(H_2),0.1), ~\edDelOpr(H_1,|,0,1) \rangle.$\\
\normalsize
This inverse mapping,   applied on  the LTG of Figure~\ref{fig:hospLTG}, gives the RTG of Figure~\ref{fig:hospRTG}.

%% file: SchEvolMap-corrector.tex
In~\cite{ABS13}, given a  well-formed XML tree $t$,  a schema $G$ and a non negative threshold $th$, \dc\  finds
every tree $t'$ valid \wrt\ $G$ such that the edit distance between $t$  and $t'$ is no higher than $th$.  
Contrary to most other approaches, \cite{ABS13} considers the correction as an enumeration problem rather than
a decision problem and computes all the  possible corrections on $t$.
The algorithm, proved to be correct and complete in~\cite{ABS13},
consists in fulfilling an edit distance matrix which stores the relevant edit operation sequences allowing to obtain the corrected trees.
The theoretical exponential complexity of \dc\  is related to the fact that edit sequences and
the corresponding corrections are generated and that the correction set is complete. 

In this paper, contrary to~\cite{ABS13},  we do not consider   all the  possible corrections on $t$.
The correction of XML documents is guided by a given mapping.
 For each edit operation on $S$, to obtain $T$,  we analyse what should be the corresponding update on  document $t$.
When this update violates validity, we use \dc\ to propose corrections to the subtree involved in the update.

%% file: SchEvolMap-adaptDocument.tex
This section   outlines  our  data translation method which is guided by a schema mapping.
Our method consists in  performing a list of changes  on  XML documents, in accordance with 
 the edit operations found in the mapping. 
For example, adding or deleting a regular expression in a rule under the operator '$.$' is a mapping operation that provokes, respectively,  the insertion or the deletion of a subtree in an originally valid XML tree (to maintain its validity). 
Similarly, renaming a non-terminal $A$ by $B$, provokes the substitution of the subtree generated by $A$ into the subtree generated by $B$.
When  local correction on XML  subtrees  are needed, \dc\ is  used  to ensure document validity. 
%

Consider an XML tree $t$  valid \wrt\ schema $S$ and a mapping $m$ from $S$ to $T$.
Our method can be summarized in two steps:
\begin{enumerate}
  \item \label{step1} Since $t$  belongs to the language  $L(S)$, it is possible to associate a non-terminal $A$ with each tree node position $p$ generated by this non-terminal.
  We analyse $t$, detect each   non-terminal and   annotate it  with its corresponding position $u$  in the  used production rule.
  This annotation respects the format  $(p, A^u)$.
  For example,  in Figure~\ref{fig:Docs}(b), we notice that the tree node \textit{bill} is generated  by the non-terminal $B$ whose position in $t^r_{I_1}$ is
  $0.2$, noted as   $(1.0,B^{0.2})$.

  \item  Each edit operation $ed$ in  $m$  activates a set of modifications on $t$.
 When   $ed$  transforms a grammar into a new grammar containing the previous one,  the set of modifications is empty.
 Otherwise, our method consists in traversing $t$  (marked as in step~\ref{step1})
  in order to find the tree positions which may be  affected due to $ed$.
  Modifications on $t$ are defined according to each edit operation and are not detailed here due to the lack of space. 
  Obviously,  if no position is affected,  $t$ does not change.
  
\end{enumerate}

\vspace{-0.2cm}
\noindent
The inverse mapping (Section~\ref{SchEvolMap-compInv}) guides changes on tree $t_1$ (Figure~\ref{fig:Docs}(b)).
\vspace{-0.2cm}
\begin{enumerate}

  \item Nodes in $t_1$ are  annotated  in blue, Figure~\ref{fig:Docs}(b).
  \item $\edInsTreeRule(I_3,info,B)$ implies zero change on $t_1$. By inserting a new production rule in our grammar in Figure~\ref{fig:hospLTG}, we obtain a new grammar
        which contains the previous one. The new grammar generates also $t_1$.
  \item $\edRelElm(H_1,I_1,I_3,0.2.0)$ requires  each  $I_1$   s.t.  $t^r_{H_1}(0.2.0)= I_1$  to be renamed  $I_3$.
         As  in $t_1$ (Figure~\ref{fig:Docs}(b)), there is no annotation where  $I_1^{0.2.0}$ is  a child of annotation  $(\epsilon, H_1)$, no changes are performed on $t_1$.
  \item $\edDelTree(I_1,B,0.2)$ requires to delete $B$ s.t. $t^r_{I_1}(0.2)= B$.  
  As in Figure~\ref{fig:Docs}(b) annotation $(1.0, B^{0.2})$ (node labeled \textit{bill})  is a child of annotation  $(1, I_1^{0.0.0})$ changes should be performed on $t_1$.
  Since    we delete  $B$    from the expression $(P \mid T) \mid (C \mid Pol) \mid B$ we must replace the subtree generated by $B$ by a subtree generated by 
        $(P \mid T) \mid (C \mid Pol)$ for preserving the validity of $t_1$ \wrt\   the new grammar. 
        For doing that,   we launch \dc\ on subtree at position $1.0$ in $t_1$ and the expression $(P \mid T) \mid (C \mid Pol)$
         that, in this case,   computes only  the  subtree \textit{patient(SSN,name,visitInfo(trId,date))}  with a (minimal) cost of $5$.
  		Let $t_2$ be the tree with the new subtree (Figure~\ref{fig:Docs}(c)).
  \item All other operations in the mapping imply zero change on $t_2$ due to the same reasons as stated in  (2) or (3). 
       As expected, the result is  in Figure~\ref{fig:Docs}(c).
\end{enumerate}

%% file: SchEvolMap-relwork.tex
Much other work deals with schema evolution.
In~\cite{FKPT11} authors show how inversion and composition models schema evolution, by expressing schema constraints in a logical formalism.
Second order logic is needed to express some mapping compositions.
This approach is the basis for proposals in \cite{LibMu09,PopHan07,PopYu05} dealing with XML schema evolution.
We believe that the use of edit operations makes our approach simpler than theirs and 
gets on well  with our previous work concerning  XML document correction.
Other proposals, such as those in~\cite{Nob13,Leo07,NosKH13,CGM11,SuzFuk08},  use edit operations. 
ELaX (Evolution Language for XML-Schema) in~\cite{NosKH13} and Exup~\cite{CGM11} are a domain-specific language that proposes to  handle modifications on  XSD and to express
such modifications formally.
Contrary to us, approaches  in~\cite{Leo07,SuzFuk08} only consider LTG evolution.
In~\cite{Nob13} we find a proposal that is closer to ours, dealing with RTG evolution.
An important originality of our approach is the \textit{automatic} generation of a \textit{conservative extension} of an RTG into an LTG, following the lines of~\cite{CHMR13}.
The use of a schema mapping to guide document adaptation is also considered in~\cite{SuzFuk08,CGM11}.
However in~\cite{SuzFuk08}, when the original grammar is ambiguous allowing  more than one solution, their method fails. 
Our approach may propose different solutions to be chosen by the user.
\docT\ is  guided by a  mapping and produces documents with corrections that do not exceed a threshold.

%% file: SchEvolMap-FinalRem.tex

\esch (\cite{CHMR13})    returns a conservative extended grammar. 
\mapG\ automatically produces a mapping for this conservative type evolution. 
Having  a mapping allows  any evolution  (conservative or not) via inversion or composition.
\docT\ uses \dc\ locally and follows a given mapping to propose XML document adaptations.
Thus, all \textit{local} solutions under a given threshold are  produced.
Our system offers  flexibility. 
In  \docT\ (Section~\ref{SchEvolMap-adaptDoc}, step~\ref{step1}),   different annotations are possible
 (indeed, for  1-ambiguous LTG only one annotation is possible, but general RTG allow distinct annotations).
 Our method can produce all possible document adaptations  (\ie, those respecting  local thresholds) or let to the user the choice of following 
 just a fixed number of them.
 The user can also adapt edit operation costs  according to his priorities. 
      A prototype, implemented in Java,  is been tested.
 As a first experiment, we have produced an LTG, in $24 ms$, by merging  the grammars obtained 
 from
\textit{dblp}  DTD\footnote{\url{http://dblp.uni-trier.de/xml/dblp.dtd}} and
\textit{HAL XSD}\footnote{\url{http://import.ccsd.cnrs.fr/xsd/generationAuto.php?instance=hal}}.
  \mapG\ returned a  19-operation mapping. 
Then \docT\ was used to adapt a 52-node document valid \wrt\ the computed  LTG  toward  
the \textit{HAL} grammar, giving, in this case, 36 solutions  in $22.6~s$.
As in this  test,  all possible translations can be  considered, but
the user may also interfere in  an intermediate step, making choices before the end of the complete computation - guiding and,
thus, restricting the number of solutions.  We are currently working on a friendly interface to facilitate this intermediate interference.

Our \textit{ToolBox}  offers  schema evolution  mechanisms  accompanied  by an automatic adaptation of XML documents.
Its  conservative aspect guarantees  great flexibility when a global integrated system co-exists with local ones.